\documentstyle[12pt,iopfts]{ioplppt}

\begin{document}

\title{Coherent states for the hydrogen atom}
\author{S A Pol'shin\ftnote{1}{E-mail: itl593@online.kharkov.ua}}
\date{}

\address{Department of Physics, Kharkov National University, \\
  Svobody Sq., 4, 61077, Kharkov, Ukraine }

\jl{1}


\begin{abstract}
We construct a system of coherent states for the hydrogen atom that is
expressed in terms of elementary functions. Unlike to the previous attempts
in this direction, this system possesses the properties equivalent to the
most of those for the harmonic oscillator, with modifications due to the
character of the problem.
\end{abstract}

\section{Introduction}

In 1926 Schr\"{o}dinger constructed the superposition of states for the
harmonic oscillator, afterwards called the system of coherent states (CS).
It is parametrized by complex numbers and possesses a number of remarkable
properties:

{\leftskip30pt

\noindent $(A)$ In the configuration space it may be expressed in the close
form.

\noindent $(B)$ The evolution operator $\e^{-\i {\cal H}t}$ transforms
an arbitrary state of the system into the state also belonging to the system.

\noindent $(C)$ Each state returns to its initial value after the lapse of
the time $T=2\pi /\omega $, i.e. the operator $\e^{-\i {\cal H}T}$ maps
each state onto itself.

\noindent $(D)$ Each state of the system moves classically, i.e. the
expectation values of coordinates and momenta for an arbitrary state have
the same temporally dependence as those for the corresponding classical
problem.

\noindent $(E)$ The system of CS yields the resolution of the identity.

\noindent $(F)$ For each state of the  system the uncertainty $\Delta
p\Delta x$ attains its minimum possible value.

\noindent $(G)$ The system is invariant under the action of the
Heisenberg-Weyl group.

\noindent $(H)$ Each state of the system is well-localized in the
configuration space.

}

These properties are considered in details, for example, in the book~\cite
{Perel}. The problem of generalization of this construction to the
potentials different from the harmonic one appears naturally. One of not
numerous examples of its successful solutions is a recent construction of CS
system for the one-dimensional Morse potential~\cite{Benedict}. Similarly to
the case of the harmonic oscillator, this CS system is parametrized by the
complex numbers and the expectation values of coordinates and momenta are
expressed in terms of this parameter. This CS system obeys the conditions
$(A),(E),(H)$ and an analog of the condition $(G)$ with some solvable group
different from the Heizenberg-Weyl one.

Meanwhile, the problem of construction of CS for the hydrogen atom stated by
Schr\"{o}dinger is of significant interest on its own. In this case we
should justify the set of properties the validity of which we demand. For
example, as the symmetry group of the hydrogen atom is $SO(4,2)$~\cite{BRasm}
, it is natural to replace $(G)$ by the condition

{\leftskip30pt

\noindent $(G^{\prime })$ Invariance under the action of $SO(4,2)$ group or
some  it subgroup.

}

Validity of the properties $(B),(C),(D)$ for the harmonic oscillator is a
consequence of the fact that its energy levels are multiples of the ground
one. Then, during the time of returning the ground state to its initial
value all other states do so too. For the hydrogen atom it is correct in
some fictitious time variable rather than $t$~\cite{Gerry}. This suggests us
replacing the properties $(B),(C),(D)$ by

{\leftskip30pt

\noindent $(B^{\prime})$ Stability of the system under the evolution with
respect to the mentioned fictitious time variable.

\noindent $(C^{\prime })$ If the fictitious time variable changes on the
fixed (independent on the state) value, then all states of the system return
to their initial values.

\noindent $(D^{\prime })$ During the evolution with respect to the fictitious
time variable the expectation value of $\bi x$ circumscribes an ellipse.

}

The property $(F)$ also needs modification since the dispersions $\Delta
x\Delta p_{x}$ and $\Delta r\Delta p_{r}$ are nonminimal even for the ground
state of the hydrogen atom. Instead of $(F)$, we can introduce the following
criterion of extracting the states which are most close to the classical
ones~\cite {Perel}:

{\leftskip30pt

\noindent $(F^{\prime })$ The value of
\[
\Delta C_{2}=\langle \psi |C_{2}|\psi \rangle -g^{mn}\langle \psi
|X_{m}|\psi \rangle \langle \psi |X_{n}|\psi \rangle
\]
is minimal for all the states of our system.

}

Here $X_{m},\ g^{mn}$ and
$C_{2}=g^{mn}X_{m}X_{n}$ are the generators of symmetry group of our CS
system, Cartan tensor of this group and its Casimir operator, respectively.

Starting from the Mostowski 1977 paper~\cite{Most}, many authors proposed
various systems of states obeying different sets of the above properties.
Here we shall enumerate only exact results without pretending to
completeness. Klauder~\cite{Klauder} constructed the CS system possessing
the properties $(B),(E)$. In~\cite{Maj} it was shown that for these states
one can satisfy the property $(G^{\prime })$ for  the $SO(4)$ group.
This approach was criticied by Bellomo and Stroud~\cite{Bellomo} who showed
that the properties $(C),(D),(H)$ fail to be satisfied.

Following the general Perelomov's method~\cite{Perel}, Mostowski~\cite{Most}
constructed the CS system satisfying the property $(G^{\prime })$ for the
group $SO(4,2)$. De Prunele~\cite{Prunele} considered the properties of this
system and showed that the property $(A)$ is satisfied for circular orbits
only and the property $(H)$ fails to be satisfied. Let us point out that
this CS system is a particular case of that for the space $%
SU(N,N)/S(U(N)\otimes U(N))$ introduced by Perelomov to describe the pair
creation of bosonic particles of nonzero spin in the external field~\cite
{Perel}.

Starting from the correspondence between the three-dimensional hydrogen atom
and the four-dimensional harmonic oscillator (see also~\cite{Nouri} and
references therein), Gerry~\cite{Gerry} constructed the CS system for the
hydrogen atom as a direct product of two CS systems for the $SO(3)$ group.
For this CS system the properties $(B^{\prime }),(C^{\prime }),(D^{\prime })$
and $(G^{\prime })$ for the $SO(4)$ group are satisfied.

The mentioned correspondence naturally suggests us using the basis numerated
by ''number operators''~\cite{Gerry}. Using the coordinate realization of
this basis given in~\cite{BRasm}, in the present paper we construct the CS
system obeying the properties $(A),(B^{\prime }),(C^{\prime }),(D^{\prime
}),(F^{\prime }),(G^{\prime })$ (for the $SO(3,2)$ group) and $(H)$. In the
quasiclassical limit (i.e. for great $\langle r\rangle $) it passes into the
usual plane wave, as it should be for the potential tending to zero at
infinity.

\section{Construction}

It is well known~\cite{BRasm} that the wave function of hydrogen atom in the
parabolic coordinates
\[
x+iy=\xi \eta \e^{i\phi }\qquad z=\frac{1}{2}(\xi ^{2}-\eta ^{2})\qquad r=
\frac{1}{2}(\xi ^{2}+\eta ^{2})
\]
reads
\begin{eqnarray}
\langle {\bi x}|n_{1}n_{2}m\rangle =(-1)^{n_{1}+\frac{1}{2}(m-|m|)}\frac{\e
^{\i m\phi }}{\sqrt{\pi }}\e^{-\frac{1}{2}(\xi ^{2}+\eta ^{2})}
\nonumber \\
\times (\xi \eta )^{|m|}\left( \frac{(n_{1}+|m|)!(n_{2}+|m|)!}{n_{1}!n_{2}!
}\right) ^{-1/2}L_{n_{1}}^{|m|}(\xi ^{2})L_{n_{2}}^{|m|}(\eta ^{2}).
\nonumber
\end{eqnarray}
In comparison with equation (2.4) of~\cite{BRasm} we have redenoted $\xi
\rightarrow \xi ^{2},\ \eta \rightarrow \eta ^{2}$ and corrected a misprint
in the normalization factor.

Let us consider the states
\begin{equation}
|\lambda _{1}\lambda _{2}\rangle =c_{0}\sum\limits_{n=0}^{\infty
}\sum\limits_{m=-\infty }^{\infty }(\lambda _{1}\lambda _{2})^{\frac{1}{2}
(2n+|m|+1)}\left( \frac{\lambda _{1}}{\lambda _{2}}\right) ^{m/2}|nnm\rangle
\label{hyd3a}
\end{equation}
where $\lambda _{1},\ \lambda _{2}$ are the complex numbers and $|\lambda
_{1}\lambda _{2}|<1$. Using the formulas~\cite{Bateman2}
\begin{eqnarray}
\sum\limits_{n=0}^{\infty }\frac{n!}{\Gamma (n+\alpha +1)}L_{n}^{\alpha
}(x)L_{n}^{\alpha }(y)z^{n}  \nonumber \\
\ns\lo =(1-z)^{-1}\exp \left( -z\frac{x+y}{1-z}\right) (-xyz)^{-\alpha
/2}J_{\alpha }\left( 2\frac{(-xyz)^{1/2}}{1-z}\right) \qquad |z|<1  \nonumber
\\
\bs\sum\limits_{n=-\infty }^{\infty }t^{n}J_{n}(z) =\exp \left[
(t-t^{-1})z/2\right]  \nonumber
\end{eqnarray}
we obtain
\begin{equation}
\langle {\bi x}|\lambda _{1}\lambda _{2}\rangle =\frac{c_{0}}{\sqrt{\pi }}
\frac{({\bi u}^{2})^{1/2}}{1+{\bi u}^{2}}\exp \left( \frac{r({\bi u}
^{2}-1)+2\i {\bi u}{\bi x}}{{\bi u}^{2}+1}\right)  \label{hyd5}
\end{equation}
where $\bi u$ is the vector with components
\begin{equation}
{\bi u}=\left( \frac{\i }{2}(\lambda _{2}-\lambda _{1}),\frac{1}{2}
(\lambda _{1}+\lambda _{2}),0\right) .  \label{hyd5a}
\end{equation}
It is well known that the $SO(3)$ transformations acting in the space of
vectors $|n_{1}n_{2}m\rangle $ correspond to usual rotations in the
configuration space. Then, applying to the vector $|\lambda _{1}\lambda
_{2}\rangle $ the rotation which transforms the vector~(\ref{hyd5a})
into the arbitrary complex three-vector of the same lenght, we obtain the
resulting state as a series in vectors $|n_{1}n_{2}m\rangle $ too; however
this series shall have a much more complicated form than~(\ref{hyd3a}). Then
we shall consider $\bi u$ as an arbitrary complex three-vector obeying the
condition ${\bi u}^{2}<1$; we denote the corresponding state as $|{\bi u}
\rangle $ rather than $|\lambda _{1}\lambda _{2}\rangle $.

To represent~(\ref{hyd5}) in a more compact form, we introduce the complex
space-like unit four-vector
\[
l_{\bi u}^{\mu }=\left( \i \frac{1-{\bi u}^{2}}{1+{\bi u}^{2}},
\frac{-2{\bi u}}{1+{\bi u}^{2}}\right)
\]
then $l_{\bi u}\cdot l_{\bi u}=-1$ (analogous transformation takes place for
the CS for the $SO(4,1)$ group too~\cite{my}), and the light-like forward
four-vector
\begin{equation}
n_{\bi x}^{\mu }=(r,{\bi x})\qquad n_{\bi x}\cdot n_{\bi x}=0\qquad n_{\bi
x}^{0}\geq 0.  \label{hyd6}
\end{equation}
Then we can rewrite~(\ref{hyd5}) in the form
\[
\langle {\bi x}|{\bi u}\rangle =\frac{c_{0}}{2\sqrt{\pi }}({\bi
l}_{\bi u}^{2})^{1/2}\exp (\i l_{\bi u}\cdot n_{\bi x}).
\]
From~(\ref{hyd6}) it follows that the measure $r^{-1}dV=\frac{1}{2}d(\xi
^{2})d(\eta ^{2})d\phi $ for the scalar product of wave functions of the
hydrogen atom~\cite{BRasm} coincides with the Lorentz-invariant measure over
the light cone. Then it is easily seen that for finiteness of the norm of
the vector $|{\bi u}\rangle $ the inequality
\begin{equation}
w_{\bi u}\cdot w_{\bi u}=\frac{1-2{\bi u}{\bi u}^{*}+{\bi u}^{2}{\bi u}^{*2}
}{|1+{\bi u}^{2}|^{2}}>0\qquad w_{\bi u}^{\mu }=\mathop{{\rm Im}}l_{\bi
u}^{\mu }  \label{hyd7}
\end{equation}
should be satisfied. The vectors obeying this inequality compose the
symmetric space~\cite{Piat}
\[
SO(3,2)/(SO(3)\otimes SO(2))\simeq Sp(2,{\Bbb R})/U(2).
\]
This space is that of CS for the bosonic system of two degrees of freedom~
\cite{Perel}. To clarify their connection with those of the hydrogen atom,
let us introduce two mutually commuting sets of creation-destruction
operators:
\[
\lbrack a_{\alpha },a_{\beta }^{\dagger }]=[b_{\alpha },b_{\beta }^{\dagger
}]=\delta _{\alpha \beta }\qquad \alpha ,\beta =1,2
\]
such as $a_{\alpha }|0\rangle =b_{\alpha }|0\rangle =0$ at $\alpha =1,2$,
where $|0\rangle \equiv |{\bi u}={\bi 0}\rangle =|n_{1}=n_{2}=m=0\rangle .$
Then the arbitrary vector $|n_{1}n_{2}m\rangle $ may be obtained acting by
the some combination of operators $a_{\alpha }^{\dagger },b_{\alpha
}^{\dagger }$ onto the vector $|0\rangle $. Then we can define the
representation of the $SO(3,2)$ group acting in the space of vectors
$|n_{1}n_{2}m\rangle $ in the following way~\cite{BRasm}
\begin{equation}
\eqalign{ L_{ij}=\frac{1}{2}(a^\dagger \sigma_k a +b^\dagger \sigma_k b)
\qquad L_{i5}=-\frac{1}{2}(a^\dagger\sigma_i C b^\dagger -a C\sigma_i b) \\
L_{i0}=\frac{1}{2\i}(a^\dagger\sigma_i C b^\dagger + a C\sigma_i b) \qquad
L_{50}=\frac{1}{2}(a^\dagger a+b^\dagger b+2) }  \label{hyd7a}
\end{equation}
where $C=\i \sigma _{2}$. These generators obey the commutation
relations
\begin{equation}
\lbrack L_{AB},L_{CD}]=\i (\eta _{AD}L_{BC}+\eta _{BC}L_{AD}-\eta
_{AC}L_{BD}-\eta _{BD}L_{AC})  \label{comm}
\end{equation}
where $A,B,\ldots =0,\ldots ,3,5$ and $\eta _{AB}=(+1,-1,-1,-1,+1)$. In
comparison with the notations of Barut and Rasmussen~\cite{BRasm} we
supressed the fourth coordinate, and the sixth coordinate is traded place
with the zero one. Let us introduce the new set of operators
\begin{eqnarray}
A_{\alpha } =\frac{1}{\sqrt{2}}(a_{\alpha }+b_{\alpha })\qquad B_{\alpha }=
\frac{1}{\sqrt{2}}(a_{\alpha }-b_{\alpha })  \nonumber \\
\left[ A_{\alpha },A_{\beta }^{\dagger }\right]  =\left[ B_{\alpha
},B_{\beta }^{\dagger }\right] =\delta _{\alpha \beta }.  \nonumber
\end{eqnarray}
All other commutators vanish. Since the matrices $C\sigma _{i}$ and $\sigma
_{i}C$ are symmetric then the generators~(\ref{hyd7a}) are the linear
combination of generatores of the $Sp(2,{\Bbb R})\simeq SO(3,2)$ group
\[
X_{\alpha \beta }=A_{\alpha }A_{\beta }\qquad X_{\alpha \beta }^{\dagger
}=A_{\alpha }^{\dagger }A_{\beta }^{\dagger }\qquad Y_{\alpha \beta }=
\frac{1}{2}(A_{\alpha }A_{\beta }^{\dagger }+A_{\beta }^{\dagger }A_{\alpha })
\]
and of those obtained from the above ones by replacing $A$ to $B$. Then the
$SO(3,2)$ group acts as a group of canonical $Sp(2,{\Bbb R})$ transformations
of each set $(A_{\alpha },A_{\alpha }^{\dagger })$ and $(B_{\alpha
},B_{\alpha }^{\dagger })$ separately.

Putting ${\bi w}_{\bi u}={\bi 0}$ by virtue of the Lorentz-invariance for
the normalization factor we obtain
\begin{equation}
|c_{0}|^{2}=\frac{1-2{\bi u}{\bi u}^{*}+{\bi u}^{2}{\bi u}^{*2}}{|{\bi u}
^{2}|}.  \label{hyd8}
\end{equation}
Then the normalized CS system is
\[
\langle {\bi x}|{\bi u}\rangle =\frac{1}{\pi ^{1/2}}(w_{\bi u}\cdot w_{\bi
u})^{1/2}\exp (\i l_{\bi u}\cdot n_{\bi x}).
\]

\section{Properties}

It is well known that the generator $L_{50}$ possesses the property~\cite
{BRasm}
\[
L_{50}|n_{1}n_{2}m\rangle =(n_{1}+n_{2}+|m|+1)|n_{1}n_{2}m\rangle .
\]
Then, using~(\ref{hyd3a}) and~(\ref{hyd8}) we obtain
\begin{equation}
\e^{\i \varepsilon L_{50}}|{\bi u}\rangle =\e^{\i \varphi
(\varepsilon )}|{\bi u}e^{i\varepsilon }\rangle .  \label{hyd9}
\end{equation}
Then, due to the Lorentz-invariance of our CS system, it follows from the
commutation relations~(\ref{comm}) that this system is invariant under the
action of the full $SO(3,2)$ group.

The generator $L_{50}$ corresponds to evolution with respect to the
fictitious time variable~\cite{Gerry}. Then the CS system we have
constructed obeys the properties $(B^{\prime }),(C^{\prime })$.

Let us consider the spatial distribution of the probability density of our
CS. Denoting
\[
w_{\bi u}^{\perp }=\left[ (w_{\bi u}^{1})^{2}+(w_{\bi u}^{2})^{2}\right]
^{1/2}\qquad w_{\bi u}^{1}=w_{\bi u}^{\perp }\cos \alpha _{\bi u}\qquad
w_{\bi u}^{2}=w_{\bi u}^{\perp }\sin \alpha _{\bi u}
\]
we obtain
\begin{eqnarray}
|\langle {\bi x}|{\bi u}\rangle | =\frac{1}{\pi ^{1/2}}(w\cdot w)^{1/2}
\nonumber \\
\times \exp \left[ -(w_{\bi u}^{0}-w_{\bi u}^{3})\xi ^{2}-(w_{\bi
u}^{0}+w_{\bi u}^{3})\eta ^{2}+2\xi \eta w_{\bi u}^{\perp }\cos (\phi
-\alpha _{\bi u})\right] .  \nonumber
\end{eqnarray}
It is Gaussian with respect to the variables $\xi $ and $\eta $ separately
and then the property $(H)$ is satisfied. Using the Lorentz-invariance, it is
easy to show that the equalities
\begin{equation}
\eqalign{ \langle {\bi u}|n_{\bi x}^\mu |{\bi u}\rangle= \frac{w_{\bi
u}^\mu}{w_{\bi u} \cdot w_{\bi u}} \nonumber \\ \langle {\bi u}|n_{\bi
x}^\mu n_{\bi x}^\nu |{\bi u}\rangle= \frac{4w_{\bi u}^\mu w_{\bi u}^\nu
-\eta^{\mu\nu} (w_{\bi u} \cdot w_{\bi u})} {2(w_{\bi u} \cdot w_{\bi u})} }
\label{hyd10}
\end{equation}
hold. We define the expectation value of the variable $f$ as
\[
\langle f\rangle =\frac{\langle {\bi u}|rf|{\bi u}\rangle }{\langle {\bi u}
|r|{\bi u}\rangle }.
\]
Here and in~(\ref{hyd10}) we take the scalar product with the measure $
r^{-1}dV$. Without loss of generality we can consider ${\bi u}=({\bi k}
+\i {\bi m})\e^{\i \theta }$, where ${\bi k},{\bi m}\in {\Bbb R}^{3}$
and ${\bi k}{\bi m}=0$. Then using~(\ref{hyd10}) and~(\ref{hyd7}) we obtain
\[
\langle {\bi x}\rangle =\frac{2{\bi w}_{\bi u}}{w_{\bi u}\cdot w_{\bi u}}=-4
\frac{(1+{\bi k}^{2}-{\bi m}^{2}){\bi m}\cos \theta +(1+{\bi m}^{2}-{\bi k}
^{2}){\bi k}\sin \theta }{1-2({\bi k}^{2}+{\bi m}^{2})+({\bi k}^{2}-{\bi m}
^{2})^{2}}.
\]
In view of~(\ref{hyd9}) from the above expression the property $(D^{\prime
}) $ follows immediately. Let us emphasize that unlike the case of harmonic
oscillator, changing $\langle {\bi x}\rangle $ does not mean changing the
position of the probability density maximum. With the arbitrary $\bi u$ this
maximum is situated at the point ${\bi x}={\bi 0}$ -- at the center of the
ellipse. This is a result of the fact that for the arbitrary $\bi u$ the
states with $n_{1}=n_{2}=0$ dominate.

For our CS system the property $(F^{\prime })$ is satisfied. Indeed, we can
consider our CS system as that constructed using the general Perelomov's
method~\cite{Perel} by acting the $SO(3,2)$-transformations onto the
fiducial vector $|0\rangle $ since this vector has the stationary subgroup
$SO(3)\otimes SO(2)$. Let us consider the stationary (up to multiplication by
the real constant) subalgebra ${\cal B}$ of this vector in the complexified
Lie algebra ${\cal G}^{c}$ of the $SO(3,2)$ group. The subalgebra ${\cal B}$
is composed by the generators $L_{ij},L_{i5}+\i L_{i6}$ and $L_{56}$;
together with its conjugated subalgebra $\overline{{\cal B}}$ the subalgebra
${\cal B}$ exhausts the full algebra ${\cal G}^{c}$ i.e. the subalgebra $%
{\cal B}$ possesses the so-called maximality property (in the case of full
conformal group this was pointed out in~\cite{Prunele}). From the other hand,
for an arbitrary Lie group the property $(F^{\prime })$ is satisfied if we
construct our CS system starting from the fiducial vector which has the
maximal stationary subalgebra in the Lie algebra ${\cal G}^{c}$~\cite{Perel}.

It is well known that the Shilov boundary of the space $Sp(2,{\Bbb R})/U(2)$
is $S^{1}\times S^{2}$~\cite{Piat}; the passage to it may be performed
putting ${\bi u}\rightarrow {\bi q}\e^{\i \beta }$, where $\bi q$ is
real and ${\bi q}^{2}=1$. Then it is readily seen that $w_{\bi u}^{\mu
}\rightarrow 0$ and from~(\ref{hyd10}) we obtain
\[
\langle r\rangle =\frac{2w_{\bi u}^{0}}{w_{\bi u}\cdot w_{\bi u}}\rightarrow
\infty .
\]
Then passage to the Shilov boundary corresponds to the quasiclassical limit.
In such a case the particle motion should become free; indeed, putting
$c_{0}=1$ and $\beta =0$ we obtain $|{\bi u}\rangle \rightarrow \e^{\i
{\bi q}{\bi x}}$ i.e. the plane wave for a particle of unit mass.

\ack

I am grateful to Yu.P.Stepanovsky for helpful discussions and for placing my
attention on the reference~\cite{BRasm}, and to A.A.Zheltukhin sending me
copy of the paper~\cite{Nouri}.

\end{document}